\documentclass{article}
\usepackage{spconf,amsmath,graphicx}
\usepackage{multirow}
\usepackage{balance}

\title{A NEW PARALLEL MESSAGE-DISTRIBUTION TECHNIQUE FOR COST-BASED STEGANOGRAPHY}
\name{Mehdi Sharifzadeh, Chirag Agarwal, Mahdi Salarian, Dan Schonfeld}
\address{University of Illinois at Chicago\\
Electrical and Computer Engineering Department\\
Chicago, Illinois 60607, United States}
\usepackage{mathtools}
\usepackage{amsmath}
\usepackage{amssymb}
\usepackage{cite}
\DeclareMathOperator*{\argmin} {arg\,min}

\DeclareMathOperator*{\E} {E}
\DeclareMathOperator*{\entropy} {H}
\DeclarePairedDelimiter\abs{\lvert}{\rvert}%

\begin{document}
%
\maketitle
\begin{abstract}
This paper presents two novel approaches to increase performance bounds of image steganography under the criteria of minimizing distortion. First, in order to efficiently use the images' capacities, we propose using parallel images in the embedding stage\footnote{Our work has been done independently of Ker, Andrew David, and Tomas Pevny. ``Batch steganography in the real world." Proceedings of the on Multimedia and security. ACM, 2012., with a different approach and formulation.}. The result is then used to prove sub-optimality of the message distribution technique used by all cost based algorithms including HUGO, S-UNIWARD, and HILL. Second, a new distribution approach is presented to further improve the security of these algorithms. Experiments show that this distribution method avoids embedding in smooth regions and thus achieves a better performance, measured by state-of-the-art steganalysis, when compared with the current used distribution.
\end{abstract}
\section{Introduction}
\label{sec:intro}
The steganography problem was modeled by the prisoner’s problem where Alice and Bob want to communicate without raising any suspicion from the warden, Wendy \cite{prisoner}. In this problem, a hidden message is embedded by Alice in a cover medium with a private or public key, producing a stego medium from which the message should be decodable by Bob. The process of transferring the stego message is monitored by Wendy, who can also alter it to prevent any covert communication \cite{covert} between prisoners in case of being an active warden. However, we are only considering the case of passive warden, who only examines the media in transmission and try to reveal the existence of any hidden message. Security of a steganography method is measured by how difficult the disclosing is for the warden or the steganalyzer, which is formally formulated in \cite{cachin1998information}, where Cachin defines perfectly secure and $\epsilon$-secure steganography. This problem was investigated further from information theoretic point of view by Moulin et al. \cite{moulin2003information,moulin2007capacity}, where theoretic bounds were calculated for both cases. It has been shown that cover generation method can reach these bounds \cite{ryabko2009asymptotically}. However, these approaches need accurate knowledge of the probability distribution of the cover media, which is unfeasible for empirical non-stationary media such as images or videos.

In practical approaches for steganography, Alice tries to embed her message while minimizing the caused distortion which can be formulated to a source coding problem with a fidelity criterion \cite{shannon1959coding}. One approach for minimizing the distortion is to make very small changes in spatial domain at the noise level. For example, one of the most popular image steganography methods is altering the least significant bit of pixels individually according to the hidden message bits \cite{cheddad2010digital,johnson1998exploring}. However, because of the dependent noise and pixel to pixel dependencies in images, it can be easily detected \cite{fridrich2001reliable}. So for achieving a better security \cite{khaleghi2015ic}, embedding should be done in more complex texture or noisy areas where noises are independent rather than smooth regions. 

This has led to a group of spatial image steganography methods that we call cost based algorithms. These approaches have two main steps, first is calculating the cost of embedding in each pixel using a suitably defined distortion function, and second is embedding the message according to the costs. Second step is solved for a rather general class of distortion functions using syndrome trellis codes \cite{filler2011minimizing,filler2010gibbs}. As a result, the main focus in cost based image steganography is on the first step, deriving a cost function; however, we will show in Section \ref{ssec:flaws} that the solution for the second step used a methodology for distributing the message among pixels from \cite{fridrich2007practical} which is not optimal. Thus, the security of all the cost based algorithms can be improved using the proposed method in Section \ref{sec:New}. In addition, they can also be improved by using multiple images in parallel, instead of embedding in individual images, which we show in Section \ref{sec:parallel}.

Examples of cost based algorithms are HUGO \cite{HUGO}, S-UNIWARD \cite{S-UNIWARD} and HILL \cite{HILL}. They utilize the same algorithm for message distribution. HUGO defines the distortion as a weighted sum of difference between SPAM feature vector of a cover image and its stego version \cite{pevny2010steganalysis}. In spatial universal wavelet relative distortion (S-UNIWARD) the embedding distortion is calculated using directional filter banks. HILL uses a high-pass filter to find noisy parts in an image, and then uses two low-pass filters to smooth the calculated costs.

This paper is organized as follows. We introduce notations and review the preliminaries and the problem formulation in Section \ref{sec:prior}. In Section \ref{sec:parallel}, we present a new embedding technique using parallel images. The main contribution of this paper, a new message distribution method is presented in \ref{sec:New}. Results of comparative experiments are given in Section \ref{sec:experiments} to demonstrate the effectiveness of the proposed distribution method. Conclusions are drawn in Section \ref{sec:conclusions}.

\section{Cost-Based Steganography}
\label{sec:prior}

\subsection{Notation}
\label{ssec:notation}

Capital and lower case bold face symbols are used for matrices and vectors respectively. $\entropy$ is the entropy function in bits, and all the logarithms are binary. $\E$ is the probability expectation function.

\subsection{Problem Formulation}
\label{ssec:formulation}

We use $\textbf{X}=(x_{ij})^{n_1 \times n_2} \in \mathcal{X}=\{0,\dots,255\}^{n_1 \times n_2}$, set of all 8-bit gray-scale images of size $n_1 \times n_2$,  for cover image. Assuming a ternary embedding scenario, $\textbf{Y}=(y_{ij})^{n_1 \times n_2}$ is the stego image, where $y_{ij} \in \mathcal{I}_{ij}=\{\max(0,x_{ij}-1),x_{ij},\min(255,x_{ij}+1)\}$. As a result, the embedding pattern, defined as $\textbf{S}=\textbf{Y}-\textbf{X}=(s_{ij})^{n_1 \times n_2} \in \mathcal{S}=\{-1,0,+1\}^{n_1 \times n_2}$, is the coded stego message using practical coding schemes as syndrome trellis codes \cite{filler2011minimizing}, and it is chosen according to probability distribution $p(\textbf{S})$. 

In case of having a fixed relative payload, total number of bits over the total number of pixels, the goal is first to define a distance function $D( \textbf{X}, \textbf{Y}): \mathcal{X} \times \mathcal{X} \to \mathbb{R}$ for calculating embedding impact, then to solve the constrained optimization problem below using the suitably defined function $D$. 

\begin{equation} \label{main}
\argmin_{p} \E(D(\textbf{X},\textbf{Y})) = \argmin_{p} \sum_{\textbf{S} \in \mathcal{S}}{D(\textbf{X},\textbf{Y})\times p(\textbf{S})}
\end{equation}
\begin{equation} \label{message constraint}
m = \entropy(p) \triangleq -\sum_{\textbf{S} \in \mathcal{S}}{p(\textbf{S})\times \log\big(p(\textbf{S})\big)}
\end{equation}
where $m$ is the length of the message in bits.

Below is the solution of this problem using Lagrangian multipliers method \cite{fridrich2007practical}:

\begin{equation} \label{exponential}
p(\textbf{S}) = \frac{e^{-\lambda D(\textbf{S})}}{\sum_{\textbf{S} \in \mathcal{S}}{e^{-\lambda D(\textbf{S})}}}
\end{equation}
where $\lambda$ is the Lagrangian multiplier which can be determined from (\ref{message constraint}). This solution is proved to be the optimal solution in \cite{fridrich2007practical} but our proposed algorithm shows that it is sub-optimal.

By assuming mutually independent and symmetrical embedding impacts \cite{saturatedpixels} under an additive distortion scenario, which is a reasonable assumption also made in HUGO, HILL and S-UNIWARD, distance function can be written as below:
\begin{equation} \label{additive distortion}
D(\textbf{X},\textbf{Y}) = \sum\limits_{i=1}^{n_1} \sum\limits_{j=1}^{n_2}
 \rho_{ij} \abs{x_{ij}-y_{ij}}
\end{equation}
where $\rho_{ij}$ is the embedding impact of changing only one pixel by $\pm1$ and it can also be called as the cost of embedding. This will result in the following probability distribution for embedding changes:
\begin{equation} \label{additive probability}
p(s_{ij}) = 
  \begin{cases} 
   \frac{e^{-\lambda \rho_{ij} }}{1+2e^{-\lambda  \rho_{ij}}} & \text{if  $s_{ij} = \pm 1$} \\
   \frac{1}{1+2e^{-\lambda  \rho_{ij}}}       & \text{if  $s_{ij} = 0$}
  \end{cases}
\end{equation}

A considerably large portion of the recent effort in steganography has been focused on calculating $\rho$ values, the main difference among the state-of-the-art algorithms. However, to the best of the authors' knowledge, no one has looked into a message distribution different from (\ref{additive probability}). Investigating this, makes it possible to improve security of all the cost based algorithms.
\section{Parallel Embedding}
\label{sec:parallel}
Most of the state of the art approaches for steganography use images or video frames individually for embedding data; however parallel embedding is a promising alternative. In the proposed parallel embedding method, $n$ images or video frames are grouped together and the sum of their payloads is the given amount of data to be hidden in all of them; however, the message is not distributed evenly among  them and the payload for each image depends on its capacity. By using $n$ images grouped together for embedding, more data will be stored in an image with more complex texture or noisy areas rather than an image with mostly smooth regions. As a result, the capacity of each image is used more efficiently. 

The proposed method can be applied to any steganography method. For paralleling cost based methods, the costs is calculated individually for each image; however, the embedding is done on $n$ images grouped together. This means that $\rho$ values are calculated independently for each image. But $\lambda$ is calculated once for each group, and the message is embedded using probability distribution explained in (\ref{additive probability}). We will show that using this process, the detectability of S-UNIWARD, a cost-based algorithm, decreases for some payloads.

\begin{figure}[t]
\begin{minipage}[b]{1.0\linewidth}
  \centering
  \centerline{\includegraphics[width=8.5cm]{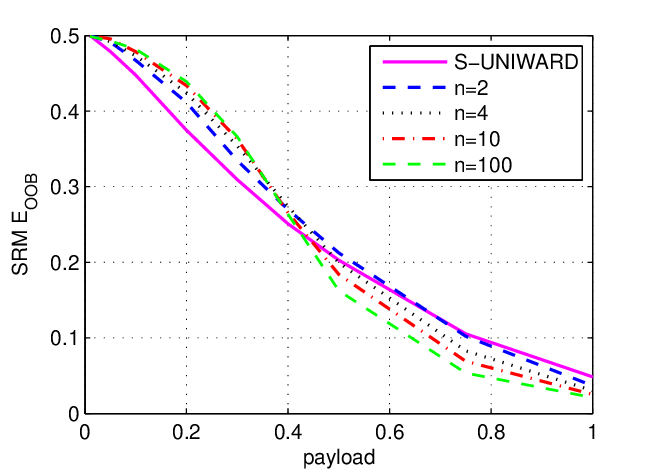}}
\end{minipage}
\caption{Out-of-Bag error (detection error) versus relative payload bits per pixel (bpp) of S-UNIWARD algorithm and its parallel versions with different $n$ values applied on BOSSbase database ver. 1.01 and measured by an ensemble classifier trained on SRM features.}
\label{figure 1}
\end{figure}

Fig. \ref{figure 1} shows the results of the experiment for comparing S-UNIWARD algorithm with its parallel versions. All the experiments are done on BOSSbase database ver. 1.01 \cite{BOSSbase}. The out-of-bag error, $E_{OOB}$, is calculated by an ensemble classifier trained and tested on SRM features \cite{SRM} extracted from images. A more detailed explanation about all the experiments and their settings is given in Section \ref{sec:experiments}.

\begin{figure}[t]
\begin{minipage}[b]{1.0\linewidth}
  \centering
  \centerline{\includegraphics[width=8.5cm]{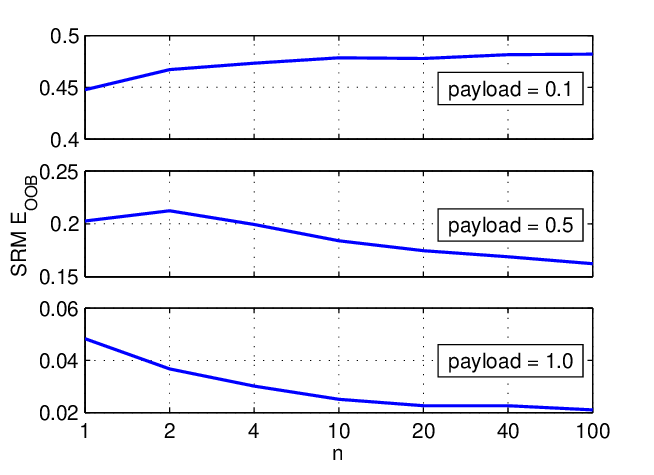}}
\end{minipage}
\caption{Out-of-Bag error (detection error) versus $n$ (number of parallel images) of parallel versions of S-UNIWARD algorithm applied on BOSSbase database ver. 1.01 and measured by an ensemble classifier trained on SRM features.}
\label{figure 2}
\end{figure}

\subsection{Flaws in Message Distribution}
\label{ssec:flaws}
The detectability should be a decreasing function of number of paralleled images if the message distribution among paralleled images is optimized. For having a better understanding of why this is true, let’s compare using $n$ and $2n$ number of parallel images for ease of explanation. Let’s assume that the average payload for both cases is $m$ bits per image. In the original algorithms $m$ bits are embedded in each image but in the first case, total number of $m \times n$ bits are embedded in $n$ images grouped together, and $m \times 2n$ bits in $2n$ images in the second case. However, in the second case we can embed $m \times n$ bits in the first $n$ images and another $m \times n$ bits in the second $n$ images which will result in the same stego images and detectability as the first case. As a result, if we optimize the way we split the message among the images, the performance should not decrease by using more images in parallel. 

Fig. \ref{figure 2} shows that there is an optimized number of parallel images for each payload which is 1 for high (near 1) payloads. This shows the message distribution over $n$ parallel images, which is the same as the message distribution among pixels of each image, is not optimized. This means the message distribution among pixels shown in (\ref{exponential}) and (\ref{additive probability}) is sub-optimal. We propose a new distributing function in the next section which results in lower detectability.

\begin{figure*}[t]
\begin{minipage}[b]{1.0\linewidth}
  \centering
  \hspace{-1.73cm}\centerline{\includegraphics[width=21.3cm]{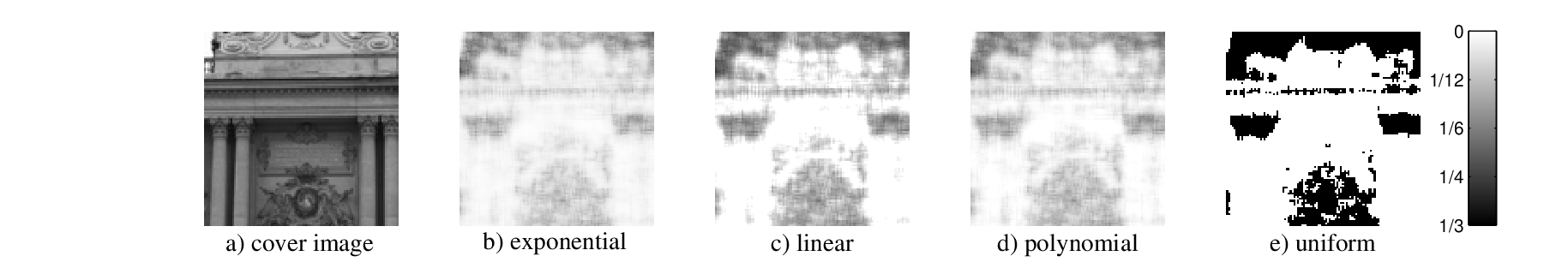}}
\end{minipage}
\vspace{-8mm}
\caption{a) Cropped version of 195.pgm, b-e) the probability of embedding $\pm1$ for 0.3 bpp calculated using S-UNIWARD with different distribution techniques. The color-map on the right hand side shows the mapping from probabilities to colors.}
\label{pics}
\vspace{-3mm}
\end{figure*}

\section{Message Distribution}
\label{sec:New}

Using the same view point as \cite{sharifzadeh2015}, water filling problem's solution can be utilized which is used in solving power allocation for Gaussian vector channels with power constraint \cite{waterfilling}. Therefore, message can be distributed with linear measure instead of the exponential measure which changes (\ref{additive probability}) to the formulation below by having the same assumption of mutually independent and symmetrical embedding impacts under an additive distortion scenario:
\begin{equation} \label{new}
p(s_{ij}) = 
  \begin{cases} 
   \max(\frac{1}{3}-\lambda \rho_{ij},0) & \text{if  $s_{ij} = \pm 1$} \\
   1-2\max(\frac{1}{3}-\lambda \rho_{ij},0)       & \text{if  $s_{ij} = 0$}
  \end{cases}
\end{equation}
where $\lambda$ is the Lagrangian multiplier which can be determined from (\ref{message constraint}). However, for the actual embedding using Syndrome Trellis codes \cite{filler2011minimizing}, a system of equations should be solved to transform all $p(s_{ij})$'s back to embedding costs using (\ref{additive probability}) same as the procedure shown in \cite{sedighi2015content}.This methodology can be applied to any cost based algorithm after calculating all the $\rho$ values, instead of using (\ref{additive probability}). This new distribution method avoids embedding in pixels with costs more than $(3\lambda)^{-1}$.
We have also run comparative experiments to check the effectiveness of the proposed distribution model which the results are presented in Table \ref{table1}. In the mentioned table, linear and exponential are the proposed and the previously used distributions respectively. The other two models which are uniform and polynomial are formulated in (\ref{uniform}) and (\ref{poly}) correspondingly.
\begin{equation} \label{uniform}
p(s_{ij}) = 
  \begin{cases} 
   \frac{1}{3}\theta(1-\lambda \rho_{ij}) & \text{if  $s_{ij} = \pm 1$} \\
   1-\frac{2}{3}\theta(1-\lambda \rho_{ij})       & \text{if  $s_{ij} = 0$}
  \end{cases}
\end{equation}
\begin{equation} \label{poly}
p(s_{ij}) = 
  \begin{cases} 
   \frac{1}{3}\max(1-\lambda \rho_{ij},0)^2 & \text{if  $s_{ij} = \pm 1$} \\
   1-\frac{2}{3}\max(1-\lambda \rho_{ij},0)^2        & \text{if  $s_{ij} = 0$}
  \end{cases}
\end{equation}
where $\theta$ is the unit step function. Fig. \ref{pics} visualizes the embedding probabilities for all these four measures. It can be observed that the proposed method embeds more in noisy areas comparing to exponential and polynomial models which helps in avoiding smooth regions more and increasing the security. However, uniform distribution only embeds in texture areas and totally avoids smooth regions which causes in a lower security comparing to the other three functions. 
\begin{table}[t]
\label{table1}
\centering
\caption{Out-of-Bag error for 0.3 bpp of different cost-based algorithms measured by an ensemble classifier trained on SRM features.}
\vspace{1mm}
\renewcommand{\arraystretch}{1.2}
\label{table1}
\begin{tabular}{|c|c|c|c|c|}
\hline
\multirow{2}{*}{Cost Function} & \multicolumn{4}{c|}{Distribution Model} \\ \cline{2-5} 
                               & linear & exp.  & uniform & polynomial \\ \hline
HILL                           & .3556  & .3512 & .2847   & .3614        \\ \hline
S-UNIWARD                      & .3291  & .3091 & .2645   & .3267        \\ \hline
HUGO                           & .2936  & .2813 & .2067   & .2074        \\ \hline
\end{tabular}
\vspace{-3.3mm}
\end{table}

In the next section, we have provided the experimental results for comparing few state of the art algorithms with their altered versions using the proposed technique.

\section{Experiments}
\label{sec:experiments}

\begin{figure}[t]
\begin{minipage}[b]{1.0\linewidth}
  \centerline{\includegraphics[width=8.5cm]{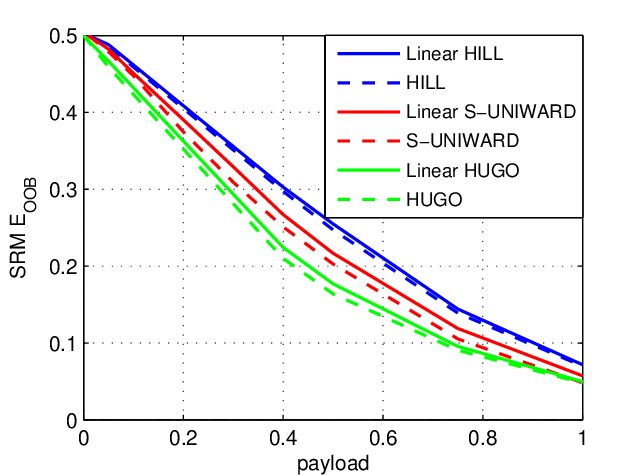}}
\end{minipage}
\vspace{-4mm}
\caption{Out-of-Bag error versus relative payload in bpp for HILL, S-UNIWARD and HUGO algorithms and their altered version using the proposed method, using SRM features.}
\label{figure 3}
\vspace{-3mm}
\end{figure}

Through out this paper, all the experiments are conducted on BOSSbase ver.1.01 database \cite{BOSSbase} containing 10,000 gray-scale $512\times512$ pixels images. 
Performance evaluations are done by an ensemble classifier steganalyzer \cite{classifier} with a 10-fold cross validation trained on 34,671 dimensional SRM feature set. 5000 and 1000 images are selected randomly in each experiment for training/validation and testing respectively. Out-of-Bag Error is reported which is the average false positive and negative rates in testing step.

HUGO is used with the same setting reported in the original paper to have its best security. S-UNIWARD algorithm is used with $\sigma=1$, shown to be optimum in \cite{S-UNIWARDsigma}. HILL algorithm is used with a $3\times3$ Ker-Bohme high-pass filter and a $3\times3$ and a $15\times15$ averaging filters as low-pass filters which is shown in the original paper to have the best security.

Fig. \ref{figure 3} shows the improvement in detectability using (\ref{new}) instead of exponential measure used by most of the cost-based steganography algorithms.
\section{Conclusions}
\label{sec:conclusions}
\vspace{-3mm}
Two methods are presented to boost the performance of every cost based image steganography algorithms. First, we showed that using parallel images will result in a better security for some payloads which is caused by using each image capacity more efficiently. Furthermore, it was shown that there is no disadvantage in using parallel images in case of having an optimized message distribution. However, the results of parallel embedding for prior works show for higher payloads it will decrease the security. Thus, we concluded that the message distribution used by all cost based algorithms is not optimal and we proposed a new distribution model. In the new model, embedding is avoided in high cost pixels in any way. We will try to prove the optimality of the new model and generalize it to non-additive cost function in our future studies.

\bibliographystyle{IEEEbib}
\bibliography{refs}
\pagebreak
\end{document}